\documentclass[journal,10pt,twoside]{IEEEtran}
\newcommand{\DRAFT}[1]{}

\usepackage{here}

\usepackage[noadjust]{cite}      

\usepackage{graphicx}  

\usepackage{subfigure} 

\usepackage{stfloats}  
\usepackage{amsmath}   

\usepackage{xspace,bbm,epic,eepic,amssymb}
\usepackage[latin1]{inputenc}
\usepackage{theorem}

\usepackage{color}

\usepackage{bbm}

\usepackage{subfigure}
\usepackage{graphicx}
\usepackage[footnotesize]{caption}
\usepackage{color}
\usepackage{cite}
\usepackage{soul}

\usepackage{algorithm}
\usepackage{algpseudocode}

\def\XS{\xspace}
\DeclareMathAlphabet{\mathb}{OML}{cmm}{b}{it}
\def\tb{{\sbm{t}}\XS}

\RequirePackage{ifthen}
\RequirePackage{calc}

\newcommand{\taille}[1][\scad]{%
\ifthenelse{#1 = -5}{}{}%
\ifthenelse{#1 = -4}{\tiny}{}%
\ifthenelse{#1 = -3}{\scriptsize}{}%
\ifthenelse{#1 = -2}{\footnotesize}{}%
\ifthenelse{#1 = -1}{\small}{}%
\ifthenelse{#1 = 0}{\normalsize}{}%
\ifthenelse{#1 = 1}{\large}{}%
\ifthenelse{#1 = 2}{\Large}{}%
\ifthenelse{#1 = 3}{\LARGE}{}%
\ifthenelse{#1 = 4}{\huge}{}%
\ifthenelse{#1 = 5}{\Huge}{}}

\newboolean{@serif}
\setboolean{@serif}{true}
\def\scad{-5} 

\newcounter{taille}
\newcommand{\sca}[2][\scad]{\setcounter{taille}{#1}%
  \ifthenelse{\boolean{@serif}}
  {{\taille[\thetaille]\textsc{#2}}}
  {\setcounter{taille}{\value{taille}-1}{\uppercase{\taille[\thetaille]#2}}}}

\def\eg{\textit{e.g.,}\XS}
\def\etal{\textit{et al.}\XS}

\def\ie{\textit{i.e.,}\XS}

\newcommand{\adresseCRAN}[1][\scad]{Campus Sciences, B.P. 70239, 
F-5   4506 Vand{\oe}uvre-l\`es-Nancy, France\XS}

   \def\stdscal#1{\langle#1\rangle}

\def\a{{\mathbf a}}

\def\x{{\mathbf x}}

\def\y{{\mathbf y}}
\def\s{{\mathbf s}}

\def\w{{\mathbf w}}

\def\r{{\mathbf r}}

\def\ta{{\tilde{\mathbf a}}}
\def\tb{{\tilde{\mathbf b}}}
\def\tc{{\tilde{\mathbf c}}}

\def\onebf{{\mathbf 1}}

\def\A{{\mathbf A}}

\def\G{{\mathbf G}}
\def\I{{\mathbf I}}

\def\P{{\mathbf P}}

\def\U{{\mathbf U}}

\def\Diag{\Lambda}

\newcommand{\Qc}{\mathcal{Q}}
\newcommand{\Qcs}{{\mathcal{Q}^\star}}
\newcommand{\Qcsc}{{\bar{\mathcal{Q}}^\star}}

\def\R{\mathbb{R}}

\def\vars_w{\sigma^2_n}
\def\vars{\sigma^2}
\def\ie{\textit{i.e.}, }
\def\eg{\textit{e.g.}, }
\def\etal{\textit{et al.} }

\def\proj{\mathbf{P}_\Qc^\bot }

\DeclareMathOperator*{\argmin}{arg\,min}
\DeclareMathOperator*{\argmax}{arg\,max}

\newtheorem{lemma}{Lemma}
\newtheorem{theorem}{Theorem}
\newtheorem{corollary}{Corollary}

\title{Relaxed Recovery Conditions for OMP/OLS \\ by Exploiting both Coherence and Decay}

\author{C.~Herzet, A.~Dr\'emeau, C.~Soussen,~\IEEEmembership{Member, IEEE}
 \thanks{C.~Herzet is with INRIA Centre Rennes - Bretagne
    Atlantique, Campus de Beaulieu, F-35042 Rennes Cedex, France
    (e-mail: Cedric.Herzet@inria.fr).}
  \thanks{A.~Dr\'emeau is with ENSTA Bretagne and Lab-STICC (UMR 6285), 42 rue Fran\c cois Verny, 29200 Brest, France (e-mail: angelique.dremeau@ensta-bretagne.fr).}    
 \thanks{C.~Soussen is with the University of Lorraine 
and CNRS at the Centre de Recherche en Automatique
de Nancy (CRAN, UMR 7039), 2 avenue de la forêt de Haye 
- TSA 60604 - F-54518 Vandoeuvre-l\`es-Nancy Cedex, France (e-mail:
    charles.soussen@univ-lorraine.fr).}
 \thanks{C. Herzet and C.~Soussen are sponsored by the French 
Agence Nationale de la Recherche (ANR), BECOSE project.
}
}

\begin{document}

\maketitle

\begin{abstract} 
  We propose extended coherence-based conditions for exact sparse
  support recovery using orthogonal matching pursuit (OMP) and
  orthogonal least squares (OLS). Unlike standard uniform guarantees,
  we embed some information about the decay of the sparse vector
  coefficients in our conditions. As a result, the standard condition
  $\mu<1/(2k-1)$ (where $\mu$ denotes the mutual coherence and $k$ the
  sparsity level) can be weakened as soon as the nonzero coefficients
  obey some decay, both in the noiseless and the bounded-noise
  scenarios. Furthermore, the resulting condition is approaching
  $\mu<1/k$ for strongly decaying sparse signals.  Finally, in the
  noiseless setting, we prove that the proposed conditions, in
  particular the bound $\mu<1/k$, are the tightest achievable
  guarantees based on mutual coherence.
\end{abstract}


\section{Introduction}
\label{sec:introduction}

In this paper, we focus on two popular instances of greedy algorithms
for sparse signal approximation from linear measurements, namely
orthogonal matching pursuit (OMP)~\cite{Pati_asilomar93} and
orthogonal least squares\footnote{The OLS algorithm is also known as
  forward selection~\cite{Miller2002Subset}, Order Recursive Matching
  Pursuit (ORMP)~\cite{Cotter1999},  pure Orthogonal Matching Pursuit (OMP) ~\cite{Foucart2013Stability} and Optimized Orthogonal Matching
  Pursuit (OOMP)~\cite{RebolloNeira2002Optimized} in the literature.}
(OLS)~\cite{Chen:1950fk, Natarajan1995Sparse}.
These two iterative procedures gradually build an estimate of the
support of a sparse representation by adding one new element to it at
each iteration, and update the sparse approximation by computing the
orthogonal projection of the data vector onto the subspace yielded by
the selected support. OMP and OLS exclusively differ in the way the
new support element is selected: OMP picks the atom leading to the
maximum (absolute) correlation with the current residual while OLS
selects the atom minimizing the $\ell_2$-norm of the new residual. In
the rest of the paper, we will use the generic acronym Oxx to refer to
both OMP and OLS in all the statements that are valid for the two
procedures. 
 
 In the recent years, many researchers have studied 
conditions under which Oxx succeeds in recovering the true sparse
vector. 
 A popular approach to address this question relies on the
derivation of \textit{uniform} guarantees; the latter ensure the
success of Oxx for a given sparsity level (or a given support)
\emph{irrespective} of the magnitude of the nonzero coefficients.
This type of analyses was carried out for OMP in
\cite{Tropp2004Greed,Davenport2010Analysis} and also adapted to
several extensions of OMP in
\cite{Needell:2009fk,Donoho2012Sparse,Foucart2013Stability}. Although
OLS has been known in the literature for a few decades (under
different names \cite{Blumensath2007Difference}), uniform exact
recovery analyses of OLS have only appeared very recently,
see~\cite{Foucart2013Stability, Soussen2013Joint,Herzet2013}.

On the one hand, uniform conditions are usually quite pessimistic
since they cannot be satisfied as soon as Oxx fails for \textit{one}
particular sparse vector. As a matter of fact, it is now acknowledged
that uniform conditions typically fail in properly characterizing the
\textit{average} behavior of the considered algorithm
\cite{Eldar10,Davies12}.  In particular, in \cite{Sturm2011Sparse} the author emphasized  that the empirical behavior of OMP is much dependent on the distribution defining the  amplitudes of the nonzero coefficients. On the other hand, probabilistic analyses\footnote{We
  are referring here to analyses characterizing the success of Oxx
  with ``high probability", for a deterministic dictionary and a
  random sparse vector. This is in contrast with the probabilistic
  analyses performed in \cite{Tropp2007Signal,Fletcher2009Orthogonal},
  which focus on the \textit{uniform} success of OMP for random
  dictionaries (and a given support).} are usually quite involved to
carry out for deterministic dictionaries because of the intricate
nature of the recursions defining Oxx. It is noticeable that a probabilistic
analysis of OMP has been proposed within the multiple measurement
setup (\ie when several data vectors having a common sparsity profile
are to be simultaneously decomposed in the same
dictionary)~\cite{Gribonval08b}. In this context, the uniform recovery
guarantees can be significantly weakened within a probabilistic
framework. Nevertheless,  this result was shown to be
only relevant when the number of measurement vectors is of the same
order as the sparsity level and does therefore not apply to the single
measurement case.

In this paper, we adopt a deterministic analysis
  technique for sparse vectors whose nonzero coefficients obey some
decay. Our analysis is therefore no longer uniform since it restricts
the success of Oxx to a certain class of sparse signals. To some
extent, it may also provide insights into the success of Oxx
for random input vectors as long as one can characterize the decay of
``typical" realizations of the latter. 
 From another point of view, let us mention that a number of authors empirically observed (and then conjectured) that the worst-case scenario for Oxx corresponds to the situation where all the nonzero coefficients have the same amplitude, see \eg \cite{Dai2009Subspace,Maleki2010Optimally,Blanchard2015Conjugate,Sturm2011Sparse}. The analysis of Oxx with decaying vectors is thus also expected to bring some answer to this question since vectors with equal nonzero coefficients correspond to the particular case of ``no decay''.

 Although sparse vectors with decaying nonzero coefficients can be observed in many applications (see \cite{Mallat2008Wavelet} and \cite{Vincent2014From} for examples in the field of image and audio processing, respectively),  
 we are only aware of a few
works analyzing the success of Oxx in such a setup~\cite{Jin2008Performance,Davenport2010Analysis,Ding2012Performance,Ehler2014Quasilinear}. 
 In \cite{Jin2008Performance} the authors adopted an information-theoretical point of view: 
they derived ``rates'' (\ie dictionary dimensions and sparsity levels) under which a ``successive interference canceller'' (which can be understood as an idealized version of Oxx) can asymptotically succeed. 
 In particular, they showed that the achievable rates depend on the amplitudes of the nonzero coefficients (which, in their theoretical framework, must be known to the receiver), and thus on the decay. 
 However, their results only apply to randomly-generated dictionaries. 
In \cite{Davenport2010Analysis}, the authors provided an analysis of OMP
in terms of restricted isometry constants (RIC) and showed that the
success of OMP can be ensured by considering sufficiently decaying
vectors.  In \cite{Ding2012Performance}, Ding \etal extended this type
of result to the case of observation models corrupted by noise. 
 Finally, Ehler \etal carried out the same kind of RIC-based analysis  for some non-linear generalization of OLS in \cite{Ehler2014Quasilinear}.

In the sequel, we propose novel conditions of success  in $k$ steps for both
OMP and OLS in terms of mutual coherence of the dictionary ($k$ denotes the number of nonzero coefficients in the sparse vector).  We note
that, as long as the success of OMP and OLS  in $k$ steps is concerned\footnote{This is in
  contrast with Basis Pursuit for which RICs usually lead to more favorable conditions.}, mutual
coherence and RICs are dictionary features which offer
different perspectives on the success of Oxx: as shown in \cite[Examples 2 and 3]{Herzet2013}, 
 there are instances of
dictionaries for which the uniform mutual coherence
condition $\mu<1/(2k-1)$ is satisfied but the best-known uniform
RIC conditions \cite{Mo2012Remark,Chang2014Improved} are not, and vice versa.  The conditions derived in
this paper relax several conditions previously proposed in the literature,
and encompass them as particular cases.

We will consider a unified definition of Oxx based on the orthogonal projection of
the dictionary atoms onto the orthogonal complement of the subspace
spanned by the selected atoms, see
  \eg \cite{Soussen2013Joint}. This definition allows us to
 define both algorithms in a unifying framework and to carry
out a parallel analysis of both OMP and OLS. Our
derivations are then based on a fine analysis of the correlation
between the projected atoms involved in the iterations of Oxx.  Unlike
previous works, we highlight that the decay conditions can be relaxed
as the iterations of Oxx progress. Moreover, our guarantees are tight:
these are the best achievable coherence-based guarantees exploiting
the decay between successive ordered coefficients in the noiseless
setup.

The rest of the paper is organized as follows. Our main results are stated in section \ref{sec:Results} 
 together with some  relevant connections with the state of the
 art. The technical proofs of the results are reported in section \ref{sec:td}.


\begin{algorithm}[t]
\begin{algorithmic}[0]
\State \textbf{inputs}: $\y$, $\A$
\State \textbf{init}: $\Qc=\emptyset$
\While{$\mathrm{Card}(\Qc)< k$}
	\State $\Qc = \Qc \cup \{j\}$\vspace{0.1cm}
	\State where \vspace{0.3cm}
 	\State $\qquad$$\begin{aligned} j \in \left\{
	\begin{array}{ll}
		\argmax_{i\notin\Qc}|\langle \a_i, \r^\Qc \rangle | &\mbox{(OMP)}\\ 
		\argmin_{i\notin\Qc} \|\r^{\Qc\cup\{i\}}\|_{2} &\mbox{(OLS)} 
	\end{array} \right.
	\end{aligned}
	$ \vspace{0.1cm}
	\State 
	\State and $\r^\Qc$ is the data residual associated to active set $\Qc$
\EndWhile
\State \textbf{ouput}: $\Qc$ with $\mathrm{Card}(\Qc)=k$
\end{algorithmic}
\caption{Oxx in $k$ steps \label{alg:Oxx}}
\end{algorithm}

\section{Context and Main Results}\label{sec:Results}

Let $\y\in\R^m$  obey the following model:
\begin{align}\label{eq:defy}
\y = \A \x + \w,
\end{align}
where $\A\in\R^{m\times n}$ is a known dictionary, $\x\in\R^{n}$ is an
unknown vector and $\w\in\R^{m}$ is some additive noise with
$\|\w\|_2\leq \epsilon$. The columns $\a_i$ of the dictionary are
supposed to be normalized: $\|\a_i\|_2=1$.  We investigate conditions
ensuring that Oxx selects a subset $\Qcs$ of $k$ dictionary atoms,
where $\Qcs\subseteq\{1,\ldots,n\}$ matches the support of the $k$
largest elements of $\x$.  We have summarized the main recursions of Oxx in Algorithm \ref{alg:Oxx}. $\langle \cdot, \cdot  \rangle$ denotes the vector inner product and $\r^\Qc$ is the projection of $\y$ onto the space orthogonal to the columns of $\A$ indexed by $\Qc$. 
We refer the reader to section \ref{sec:Oxx} for a more detailed description of Oxx. 

Our derivations are based on the so-called ``$k$-step" analysis of
Oxx: Oxx will be assumed to fail as soon as \emph{one} wrong atom
$i\notin\Qcs$ is included to the estimated
support~\cite{Tropp2004Greed,Davenport2010Analysis,Soussen2013Joint}.
On the contrary, Oxx succeeds if and only if the atoms in $\Qcs$ are
selected during the first $k$ iterations.  Alternative definitions of
exact sparse recovery may be considered. In
\cite{Foucart2013Stability, Zhang2011Sparse}, the authors focused on
``delayed recovery", where Oxx is assumed to succeed if the selected
atoms contain the correct support, with possible false atom
selections. This approach will not be pursued hereafter.

Several scenarios are considered. In sections \ref{ssec:noiseless} and
\ref{sec:partialsuccesnoiseless}, we address the case where the
observation model is noiseless ($\epsilon=0$) and $\x$ is $k$-sparse
with support $\Qcs$ ($x_i\neq 0 \Leftrightarrow i\in\Qcs $).  In
section \ref{ssec:noiseless}, we focus on conditions ensuring the
recovery of $\Qcs$ from the first iteration, \ie with the initial
empty support, whereas a finer analysis at intermediate iterations is
carried out in section \ref{sec:partialsuccesnoiseless}. This analysis
allows us to provide weaker guarantees of good atom selection when
\emph{(i)} less than $k$ iterations are being performed, and when
\emph{(ii)} Oxx is known to have selected good atoms in the early
iterations.
  In section \ref{ssec:partial recovery}, we address the noisy 
scenario ($\epsilon> 0$), and the case where $\x$ is
 compressible but possibly non-sparse. 
In this case, $\Qcs$ can be thought of as the  ``head'' 
  of the signal $\x$, obtained by gathering the indices of the largest amplitudes in $\x$.

Some of the results presented below share
connections. For example, the direct part of Theorem
\ref{th:mainresult2}  (section \ref{ssec:noiseless}), dealing with $k$-step recovery and noiseless
observations, can be seen as a particular case of the results
presented in Theorems \ref{th:mainresult3} and
\ref{th:noisymainresult} (sections
  \ref{sec:partialsuccesnoiseless} and \ref{ssec:partial
    recovery}). However, we chose to follow this editorial line to
keep the discussion of the results and the relation to the
current state of the art as simple as possible.

\subsection{$k$-step Analysis in the Noiseless Setup}\label{ssec:noiseless} 
%

 The first thoughtful ``$k$-step" analysis of OMP is due to Tropp in 
\cite[Th. 3.1 and Th. 3.10]{Tropp2004Greed}. He provided a
sufficient and worst-case necessary condition for the exact recovery
of any sparse vector with \emph{a given} support $\Qcs$. Moreover, he showed that the condition
\begin{align}
\mu<\frac{1}{2k-1}, \label{eq:stdmucond}
\end{align}
where $\mu \triangleq \max_{i\neq j} | \a_i^T \a_j|$ is the
\emph{mutual coherence} of $\A$, ensures the success of OMP. The
derivation of similar conditions for OLS is more recent and is due to
Soussen \etal in \cite{Soussen2013Joint,Herzet2013}.

Condition \eqref{eq:stdmucond} is uniform, that is Oxx can recover any
$k$-sparse vector \emph{irrespective} of the amplitude of the nonzero
coefficients when \eqref{eq:stdmucond} is satisfied.  On the
other hand, it was shown in \cite[Th. 3.1]{Cai2010Stable} that
\eqref{eq:stdmucond} is tight: there exist a $k$-sparse
vector $\x$ and a dictionary
$\A$ with $\mu=\frac{1}{2k-1}$ such that Oxx selects a wrong atom at
the \emph{first} iteration\footnote{Specifically, Cai \etal point out that
  there is an identifiability issue since some data vector can be decomposed
  using two $k$-sparse representations with distinct supports.}. This
shows that one cannot expect to weaken \eqref{eq:stdmucond} for the
recovery of arbitrary $k$-sparse vectors. Nevertheless, it is
noticeable that the specific sparse vector involved in the example of
\cite{Cai2010Stable} is ``flat", that is such that
\begin{align}
x_i=\mathrm{constant} \qquad \forall\, i\in\Qcs. \label{eq:defflatv}
\end{align}

%

This is not a coincidence. 
In Theorem \ref{eq:relaxation_theorem} below, we show that weaker sufficient conditions than \eqref{eq:stdmucond} can be obtained as soon as the nonzero coefficients of $\x$ obey some decay. 
\begin{theorem} \label{eq:relaxation_theorem}
If $\x$ is a $k$-sparse vector whose nonzero amplitudes are not all equal, there exists some $\mu^\star>\frac{1}{2k-1}$ such that  Oxx recovers $\Qcs$ in $k$ steps for any dictionary with $\mu<\mu^\star$. 
\end{theorem}

 Interestingly, as mentioned in the introduction, it has been stated in many pieces of research (and accepted as a ``folk knowledge'' \cite{Sturm2011Sparse}) that sparse vectors with nonzero coefficients of equal magnitude correspond to the most difficult case for many reconstruction algorithms, see \eg \cite{Dai2009Subspace,Maleki2010Optimally,Blanchard2015Conjugate}. The result in Theorem \ref{eq:relaxation_theorem} supports this  observation by stating that, as long as the satisfaction of mutual coherence conditions for exact recovery is concerned, ``flat'' vectors correspond to the worst possible case for Oxx. In particular, a condition of success more favorable than $\mu<\frac{1}{2k-1}$ always exists as soon as the coefficients of $\x$ exhibit some decay.

 Unfortunately, 
the proof of Theorem \ref{eq:relaxation_theorem}  does not provide an optimal value for
$\mu^\star$ (as a function of the rate of decay). In fact, a precise characterization of $\mu^\star$ for
general decay patterns may be a quite difficult
task. 
In the next theorem, we provide ``horizon-1" decay conditions (\ie
conditions between consecutive elements of the ordered nonzero
coefficients) ensuring that Oxx succeeds in $k$
steps. 
In our statement, we assume without loss of generality that
\begin{align}
\Qcs=\{1,2,\ldots,k\}, \label{eq:h1}
\end{align}
 and
\begin{align}
|x_1|\geq |x_2|\geq \ldots\geq |x_k|>0.\label{eq:h2}
\end{align}

\begin{theorem}\label{th:mainresult2} If
\begin{align}
\mu < \frac{1}{k}, \label{eq:cond1}
\end{align}
and 
\begin{align}
\vert x_i \vert > \frac{2\mu (k-i)}{1-i \mu} \vert x_{i+1}\vert \qquad \forall i\in\{1,\ldots,k-1\}, \label{eq:cond2}
\end{align}
then Oxx recovers $\Qcs$ in $k$ steps. 

Conversely, both conditions \eqref{eq:cond1} 
 and  \eqref{eq:cond2} are tight in the following sense:
%
\begin{itemize}
\item There exists an instance of dictionary $\A$ with
  $\mu=\frac{1}{k}$ such that for all $k$-sparse vectors $\x$
  supported by $\Qcs$, Oxx selects a wrong atom during the first $k$
  iterations.
\item For all $j\in\{1,\ldots,k-1\}$, there exists a vector $\x^{(j)}$
  and a dictionary $\A$ of mutual coherence $\mu<\frac{1}{k}$, for
  which the inequalities \eqref{eq:cond2} hold for $i \neq j$, and
  become an equality for $i=j$, and such that Oxx with $\x^{(j)}$ as
  input selects a wrong atom at the $j$-th iteration.
\end{itemize}
\end{theorem}


 Theorem \ref{th:mainresult2} encompasses the standard condition \eqref{eq:stdmucond} as a particular  case. Indeed, the decay factor appearing in the right-hand side of condition \eqref{eq:cond2} is such that 
 \begin{align}
 \frac{2\mu (k-i)}{1-i \mu}< 1, \label{eq:decaycoef}
\end{align}
as soon as
\begin{align}\nonumber
\mu&< \frac{1}{2k-i}.
\end{align}
Thus, by virtue of our convention \eqref{eq:h2},
\eqref{eq:decaycoef} implies that condition \eqref{eq:cond2} trivially
holds for any $i$ as soon as \eqref{eq:stdmucond} is
satisfied. We also note that $\frac{2\mu (k-i)}{1-i \mu}$ is a
decreasing function of $i$ for $\mu<1/k$. Hence,  the rate of decay in \eqref{eq:cond2} becomes lower as $i$ increases.


Condition \eqref{eq:cond2} can be equivalently expressed as
\begin{align}\nonumber
\mu < \mu_i^\star \quad \mbox{with $\quad \mu_i^\star = \frac{\frac{|x_i|}{|x_{i+1}|}}{2(k-i)+i \frac{|x_i|}{|x_{i+1}|}}$ },
\end{align}
$\forall\,i\in\{1,\ldots,k-1\}$. The conditions of success stated in
Theorem \ref{th:mainresult2} can thus also be rephrased as:
\begin{align}\nonumber
\mu<\mu^\star=
\min\left(\frac{1}{k},\mu^\star_1,\ldots,\mu^\star_{k-1}\right).
\end{align}
It can be seen that possible values for $\mu^\star$ range in the interval $[\frac{1}{2k-1},\frac{1}{k}]$ and depend on the decay of the nonzero coefficients of $\x$. On the one hand, the smallest value for $\mu_i^\star$  occurs when 
$|x_i|=|x_{i+1}|$, 
 in which case $\mu_i^\star=\frac{1}{2k-i}$. Hence, we recover the standard condition \eqref{eq:stdmucond} when $|x_1|=|x_{2}|$. 
 On the other hand, $\mu_i^\star>\frac{1}{k}$ (and therefore $\mu^\star=\frac{1}{k}$) as soon as $|x_i|> 2|x_{i+1}|$. This leads to the following corollary:
  \begin{corollary}\label{cor:kstepsuccess}
If $\mu<1/k$ and $|x_i|> 2 |x_{i+1}| $ $\forall\, i\in \{1, \ldots, k-1\}$, then Oxx recovers $\Qcs$ in $k$ steps.
\end{corollary}

\begin{figure}
\centering
\includegraphics{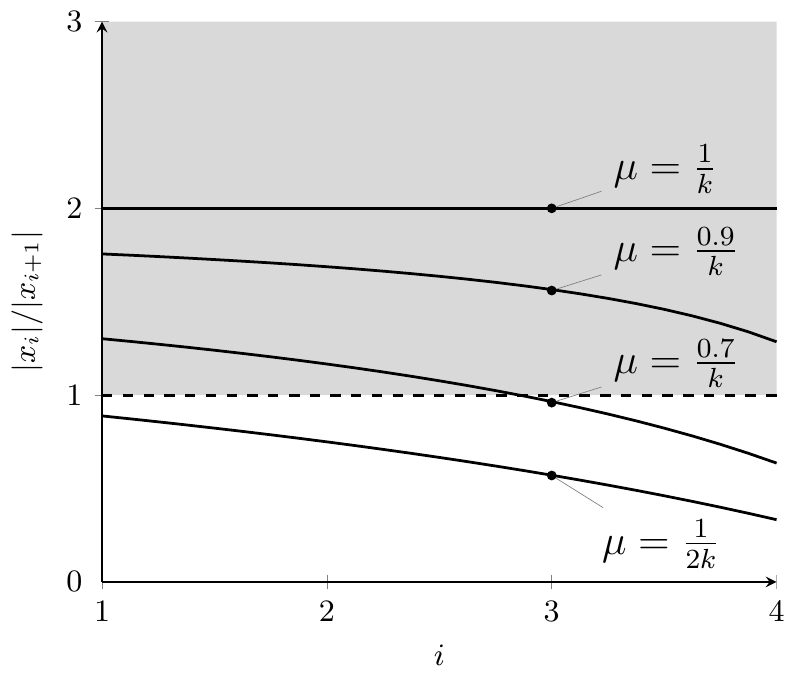}
\caption{ Graphical representation of the decay constraints of Theorem \ref{th:mainresult2} for different values of $\mu\in[\frac{1}{2k}, \frac{1}{k}]$, $k=5$. Plain curves: plot of the decay constraint $2\mu(k-i)/(1-i\mu)$ with respect to $i$. We note that because of our convention \eqref{eq:h2}, we must have $|x_i|/|x_{i+1}|\geq 1$ $\forall i$: this region corresponds to the shaded gray area. \label{fig:decay_cond}}
\end{figure}

 A graphical representation of these considerations is provided in Fig. \ref{fig:decay_cond} for $k=5$: 
the decay factor $\frac{2\mu (k-i)}{1-i \mu}$ appearing in \eqref{eq:cond2} is plotted as a function of $i$ for different values of $\mu \in [\frac{1}{2k},\frac{1}{k}]$. 
 For a given $\mu$, the region above the related curve characterizes the set of vectors $\x$ satisfying 
  the recovery conditions  of Theorem \ref{th:mainresult2}.
 We notice that the size of the region of success increases as the mutual coherence decreases.  
 In particular, when $\mu=\frac{1}{2k}<\frac{1}{2k-1}$, the curve is laying below the dashed line $|x_i|/|x_{i+1}|=1$ and \eqref{eq:cond2} is satisfied for any    $k$-sparse representation since, by convention, the nonzero entries have been sorted according to \eqref{eq:h2}. On the other side, the region of success is restricted to vectors satisfying $|x_i|/|x_{i+1}|\geq 2$ when $\mu$  is close to $\frac{1}{k}$. We note moreover that the decay constraints become less stringent as $i$ increases.  


%
%

\begin{figure} 
\centering 
\includegraphics{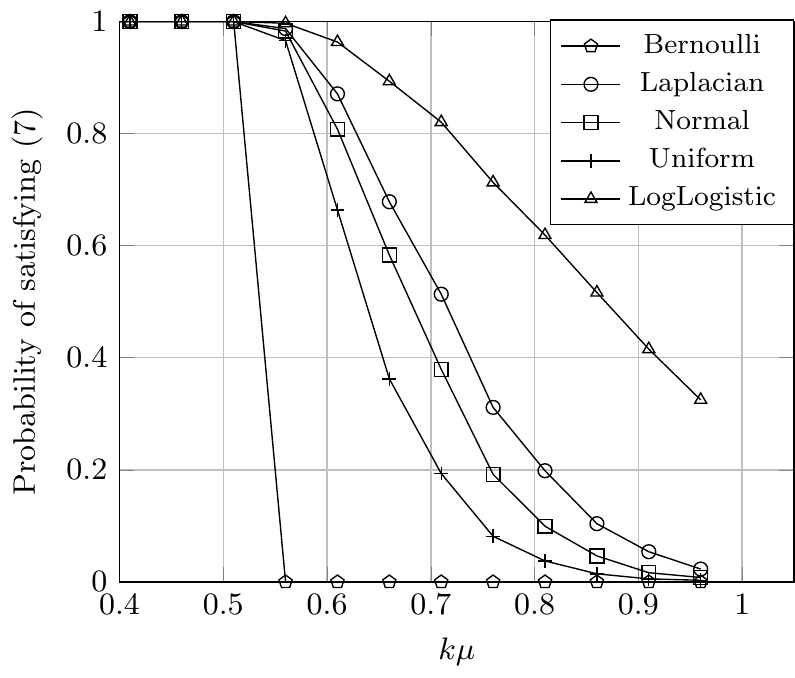}
%
%
%
 \caption{Probability of satisfying condition \eqref{eq:cond2} for random realizations drawn from different distributions versus $k\mu$.}\label{fig:empirical_prob_success}
\end{figure}

 It is also insightful to see how often nonzero coefficients drawn
 from different distributions can satisfy \eqref{eq:cond2}. In
 Fig. \ref{fig:empirical_prob_success}, we have represented the
 empirical probability that coefficients drawn from Bernoulli,
 Uniform, Normal, Laplacian and LogLogistic distributions verify the
 decay conditions of Theorem \ref{th:mainresult2}. We consider again
 the case where $k=5$ and the results are averaged over 2000
 realizations. In accordance with Theorems \ref{eq:relaxation_theorem}
 and  \ref{th:mainresult2}, the Bernoulli distribution (which always
 generates ``flat" vectors) leads to the worst results. In particular,
 conditions \eqref{eq:cond2} cannot be verified as soon as
 $k\mu\geq\frac{k}{2k-1}\simeq 0.55$. In contrast, the vectors drawn
 from the other distributions satisfy \eqref{eq:cond2} with some
 nonzero probability for any $\mu<\frac{1}{k}$ (and are therefore ensured to yield a success of Oxx). Interestingly, 
 our conclusions regarding the comparison of 
distributions is the same
 as the one observed in the empirical study of the average performance of OMP in \cite{Sturm2011Sparse}.\\

It is worth noting that not all standard sparse-representation
algorithms enjoy a relaxation of their recovery conditions when dealing with decaying
vectors. For example, the standard condition $\mu<\frac{1}{2k-1}$
cannot be improved for Basis Pursuit \cite{Chen_siam99}. Indeed, it
has been shown in \cite[Th. 3.1]{Cai2010Stable} that there exists a
dictionary $\A$ with $\mu=\frac{1}{2k-1}$ and a flat 
$k$-sparse vector $\x$ such that Basis Pursuit leads to a wrong
 support detection.\footnote{More precisely, there exists another $k$-sparse
  vector $\tilde{\x}$ such that $\|\x\|_1=\|\tilde{\x}\|_1$ and
  $\y=\A\x=\A\tilde{\x}$.}  Now, it is well-known that tight
conditions of success for Basis Pursuit only depend on the
\emph{signed support} of the sought sparse vector, see
\cite{Fuchs2004Sparse,Plumbley2007Polar}.  The existence of a
  vector $\x$ for which BP fails thus shows that BP will fail for all
  other sparse vectors with the same signed support, irrespective of
  the decay of the coefficients.

  The converse part of Theorem \ref{th:mainresult2} emphasizes that the proposed recovery conditions \eqref{eq:cond1}-\eqref{eq:cond2} are worst-case necessary in some sense. The nature of the sharpness of \eqref{eq:cond1} and \eqref{eq:cond2} is however slightly different. The tightness of \eqref{eq:cond2} is restricted to the set of ``horizon-1'' conditions, that is conditions exploiting the decay between pairs of consecutive elements in the ordered sparse vector. 
 The tightness of \eqref{eq:cond1} is of more fundamental nature since Theorem \ref{th:mainresult2}  states that there exists a dictionary such that Oxx will fail during the first $k$ iterations \textit{irrespective} of the values of the nonzero coefficients in $\x$. Hence, any mutual coherence condition ensuring $k$-step recovery and valid for general deterministic dictionaries (and in particular for the specific dictionary considered in the proof of Theorem \ref{th:mainresult2}, see section \ref{sec:proofmth}) must be of the form $\mu<\mu^\star$ with $\mu^\star\leq \frac{1}{k}$.  Said otherwise, the bound $\frac{1}{k}$ cannot be improved whatever the hypotheses made on the sparse vector. 

\subsection{Partial Recovery and Successful Termination}\label{sec:partialsuccesnoiseless}

In many applications, it is desirable to have some guarantees on the \emph{partial} success of Oxx. Two main situations may be of interest:
\begin{itemize}
\item[\emph{1})] \emph{Successful Termination:} Oxx is assumed to have selected atoms in $\Qc\subsetneq\Qcs$,
 with cardinality $\mathrm{Card}(\Qc)=g\geq 0$,  during the first $g$ iterations, and one is interested in conditions guaranteeing the selection of atoms in $\Qcs\backslash\Qc$ during the next $k-g$ iterations.
\item[\emph{2)}] \emph{Partial Support Recovery:} the focus is on
  conditions ensuring the selection of $p\leq k$ elements of
  $\Qcs$ during the  first 
$p$ iterations. 
\end{itemize}
Before we state our results, let us make a few remarks. First, the question of partial support recovery has a trivial answer in the standard ``uniform" setup. Indeed, as mentioned previously, the authors of \cite{Cai2010Stable} provided an instance of problem in which $\mu=\frac{1}{2k-1}$ and Oxx selects a wrong atom at the \emph{first} iteration. This shows that weaker coherence guarantees   
cannot be obtained for non-decaying vectors, even by restricting the
success of Oxx to partial
  support recovery. On the
contrary, we will emphasize that the paradigm of partial support recovery can be nicely addressed when accounting for the decay of the sparse vector. 

Secondly, the question of the successful termination of Oxx has
already been addressed in the uniform
setting. In \cite{Soussen2013Joint}, the authors extended Tropp's
exact recovery condition (ERC) 
to this particular setup, both for OMP and OLS. The same type of
conditions were expressed in terms of mutual coherence in
\cite[Th. 3]{Herzet2013}: if $\Qc\subsetneq\Qcs$
 is reached during the first 
$\mathrm{Card}(\Qc)=g$ iterations, then Oxx selects atoms in $\Qcs\backslash\Qc$
during the next $k-g$ iterations provided that
\begin{align}
\mu<\frac{1}{2k-g-1}. \label{eq:partialmucondition}
\end{align}
Similar to the standard $k$-step analysis \cite{Cai2010Stable}, \eqref{eq:partialmucondition} was shown to be tight: there exist a $k$-sparse vector $\x$ with support $\Qcs$, a subset $\Qc\subsetneq\Qcs$ with $\mathrm{Card}(\Qc)=g$ and a dictionary $\A$ with $\mu=\frac{1}{2k-g-1}$ such that Oxx selects atoms in $\Qc$ during the first $g$ steps and then makes a wrong decision. We show hereafter that this coherence bound can be relaxed when dealing with decaying sparse vectors.

In the statement of Theorem~\ref{th:mainresult3}, following convention \eqref{eq:h2}, we proceed to a re-ordering of the atoms 
$\a_i$ by decreasing values of their magnitudes $|x_i|$. Here, this convention is applied to the unselected atoms, which are therefore 
indexed by:
%
%
\begin{align}
\Qcs\backslash\Qc=\{1,2,\ldots,k-g\}, \label{eq:h1B}
\end{align}
 with
\begin{align}
|x_1|\geq |x_2|\geq \ldots\geq |x_{k-g}|>0.\label{eq:h2B}
\end{align}
%


 Theorem~\ref{th:mainresult3} jointly addresses both questions 
  of successful termination and partial support recovery.
\begin{theorem}\label{th:mainresult3}Assume that Oxx has selected
  $\Qc\subsetneq\Qcs$ with $\mathrm{Card}(\Qc)=g$ during the first $g$
  iterations and let $1\leq p\leq r\leq k-g$.  
\begin{itemize}
\item If
    \begin{align}
      \mu < \frac{1}{k }, \label{th:th3.a}
    \end{align}
    and the largest magnitudes of the unselected atoms after iteration $g$ satisfy
    \begin{align}
      \vert x_i \vert > \frac{2\mu (k-g-i)}{1-(g+i) \mu}\, \vert x_{i+1}\vert \qquad \forall i\in\{1,\ldots,p\}, \label{th:th3.b}
    \end{align}
    then Oxx is guaranteed to select atoms in $\Qcs\backslash\Qc$
  until all the elements in $\{1,\ldots,p\}$ have been selected.
\item If 
\begin{align}
\frac{1}{k}\leq \mu < \frac{1}{g+ r}, \label{th:th3.c}
\end{align}
 and the largest magnitudes of the unselected atoms after iteration $g$ satisfy
\begin{align}
\vert x_i \vert > \frac{2\mu (k-g-r)}{1-(g+r) \mu}\, \vert x_{i+1}\vert \qquad \forall i\in\{1,\ldots,p\}, \label{th:th3.d}
\end{align}
then Oxx is ensured to select atoms in $\Qcs\backslash\Qc$
until all the elements in $\{1,\ldots,p\}$ have 
been selected or   $g+r$ iterations have been carried out.
\end{itemize}
\end{theorem}

 Let us discuss the implications of Theorem
  \ref{th:mainresult3} on both problems of \textit{``successful
    termination"} and \textit{``partial support recovery"}. We
  specifically elaborate on the corresponding choices of $g$, $p$ and
$r$.

The paradigm of \textit{successful termination} corresponds to the
case $p=r=k-g$. In this setup, we note that conditions
\eqref{th:th3.c}-\eqref{th:th3.d} are irrelevant since
$\frac{1}{g+r}=\frac{1}{k}$,
 thus
\eqref{th:th3.c} cannot be satisfied. On the other hand, the
conditions \eqref{th:th3.a}-\eqref{th:th3.b} can be rewritten in terms of
constraints on the mutual coherence $\mu$
 involving the decay of the nonzero elements:
%
\begin{align}\nonumber
\mu<\mu^\star = \min 
\left(
\frac{1}{k }, \mu^\star_1, \ldots, \mu^\star_p \right),
\end{align}
with
\begin{align}\nonumber
\mu_i^\star = \frac{\frac{|x_i|}{|x_{i+1}|}}{2 (k-g-i)+ (g+i)\frac{|x_i|}{|x_{i+1}|}}. \nonumber
\end{align}
%
Depending on the decay of the nonzero coefficients, it can thus be
seen that $\mu^\star\in[\frac{1}{2k-g-1}, \frac{1}{k}]$. The strongest
condition corresponds to \eqref{eq:partialmucondition} and is obtained
for $|x_1|=|x_2|$; it ensures the \textit{uniform} recovery of any
$k$-sparse vector when $g$ good atoms have been selected during the
first $g$ iterations. The weakest condition, $\mu<\frac{1}{k}$, is
obtained as soon as $\frac{|x_i|}{|x_{i+1}|}\geq 2$
$\forall\,i\in\{1,\ldots,k-g\}$. We thus recover a result similar to
Corollary \ref{cor:kstepsuccess}.  Here, the decay constraints only
apply to the elements in $\Qcs\backslash\Qc$ since the elements in
$\Qc$ have already been selected by
assumption.

  Let us now discuss the particularization of Theorem
  \ref{th:mainresult3} to the problem of \textit{partial support
    recovery}   (here, $g$ is set to 0). We focus on the case where $p=
  r\leq k$. 
  %
  %
 We note that both \eqref{th:th3.a}-\eqref{th:th3.b} and 
\eqref{th:th3.c}-\eqref{th:th3.d} ensure the selection of elements of $\Qcs$ during the first $p$ iterations of Oxx 
 provided that the $p$ largest nonzero coefficients obey some ``sufficient'' decay (which is specified by either \eqref{th:th3.b} or \eqref{th:th3.d}). This leads to the following corollary for partial support  recovery:

\begin{corollary} \label{corr:partial_recovery}
Let $p\in\{1,\ldots,k\}$. If $\mu<\frac{1}{p}$ and the $p+1$ largest coefficients of $\x$ exhibit
a sufficient decay (specified by \eqref{th:th3.b} or \eqref{th:th3.d}), then Oxx selects atoms in $\Qcs$ during the first $p$ iterations. 
\end{corollary}
We note that Corollary \ref{corr:partial_recovery} 
 (which makes use of the mild assumpation $\mu<\frac{1}{p}$)
  does not guarantee that the selected atoms correspond to the $p$
  \textit{largest} coefficients of $\x$. Such guarantee can be
  obtained from \eqref{th:th3.a}-\eqref{th:th3.b} by imposing the stronger assumption $\mu<\frac{1}{k}$: 
%
%
%
\begin{corollary} \label{corr:partial_recovery2} Let $p\in\{1,\ldots,k\}$. If $\mu<\frac{1}{k}$ and the $p+1$
 largest magnitudes in $\x$ 
 exhibit a sufficient decay \eqref{th:th3.b}, then Oxx selects atoms in $\Qcs$ until the $p$
  largest components of $\x$ have been selected.
\end{corollary}

Note that Corollary \ref{corr:partial_recovery2} does not state that Oxx will select the $p$
  largest components of $\x$ during the first $p$ iterations. Their selection is however 
guaranteed during the first $k$ iterations.
 %
  %
%


\subsection{Compressible and Noisy Signals}\label{ssec:partial recovery}

In many situations, the sought vector $\x$ is not exactly $k$-sparse
but rather compressible
and possibly non-sparse. Furthermore, 
the
observations are corrupted by some additive noise
($\w\ne\mathbf{0}_m$).
%
%
We review hereafter some contributions of the literature dealing with
the success of Oxx in this particular setup and provide new strongest
results in Theorems \ref{corr:SRCnoisy} and \ref{th:noisymainresult}.
We will assume that the noise has a bounded $\ell_2$-norm,
that is $\|\w\|_2\leq\epsilon$. In the following, a signal $\x$
  will be referred to as ``$k$-compressible" as soon as the sum of the absolute values of $k$
  entries of $\x$ is large with respect to the remaining entries. Denoting by $\Qcsc\triangleq\{1,\ldots,n\}\backslash \Qcs$ the complementary subset, the $k$-compressible assumption reads $\|\x_\Qcs\|_1\gg\|\x_\Qcsc\|_1$ for some subset $\Qcs$ of
  cardinality $k$. $\x_\Qcs$ and $\x_\Qcsc$ shall be thought of as the
  head and tail of the signal $\x$, respectively. 

Let us first consider the $k$-sparse setup
($\x_\Qcsc=\mathbf{0}_{n-k}$) with noisy observations
($\epsilon>0$). Although many researchers have emphasized that the
noiseless conditions can be generalized to the case where the noise
level is low in comparison to the smallest nonzero coefficient of
$\x$ 
 \cite{Donoho2006Stable,Denis2009Greedy,BenHaim2010,Cai2011Orthogonal,Wu2013Exact},
no tight  
 condition of success for Oxx
has been proposed so far. Among the noticeable coherence-based guarantees, we can nevertheless mention the work by Donoho \etal \cite[Th. 5.1]{Donoho2006Stable} (also rediscovered in \cite[Th. 1]{Cai2011Orthogonal}), stating that OMP succeeds if 
\begin{align}
\mu &< \frac{1}{2k-1}, \label{eq:condnoisystd1}\\
\vert x_i \vert &> \frac{2 \epsilon}{1-(2k-1) \mu},  \qquad \forall i\in\{1,\ldots,k\}. \label{eq:condnoisystd2}
\end{align}

To the best of our knowledge, the extension of these results to the
success of OLS has never been made in the literature. We are neither aware of any contribution dealing with coherence-based
conditions ensuring the recovery of a particular support for
$k$-compressible vectors \emph{and} noisy observations. We address
these questions in the next theorems. The result stated in Theorem
\ref{corr:SRCnoisy} implies, as a corollary, that
\eqref{eq:condnoisystd1}-\eqref{eq:condnoisystd2} are sufficient
conditions for \emph{both} OMP and OLS. Theorem
\ref{th:noisymainresult} is an extension  of Theorem
\ref{th:mainresult2} to the noisy $k$-compressible setting. As in
subsection~\ref{ssec:noiseless}, we assume that the elements of
$\x_\Qcs$ satisfy \eqref{eq:h1}-\eqref{eq:h2}.
%
%

\begin{theorem}\label{corr:SRCnoisy}
If 
\begin{align}  
\mu < \frac{1}{2k-1}, \label{eq:condn3}
\end{align}
 and 
\begin{align}
\vert x_i \vert &> \frac{ 2 (\|\x_\Qcsc\|_1+\epsilon)}{1-(2k-i) \mu}  \qquad \forall i\in\{1,\ldots,k\}, \label{eq:condn4}
\end{align}
then Oxx selects atoms in $\Qcs$ during the first $k$ iterations.  
\end{theorem}

The conditions of Theorem \ref{corr:SRCnoisy} take the same form as
those in \eqref{eq:condnoisystd1}-\eqref{eq:condnoisystd2} but depend on 
the $\ell_1$-norm of  $\x_\Qcsc$. Moreover,   
condition \eqref{eq:condn4} 
depends on the position $i$ of the ordered
coefficients: the larger $i$,  the weaker the constraint on their amplitude. As a result, when $\|\x_\Qcsc\|_1=0$,  
Theorem \ref{corr:SRCnoisy} leads to weaker conditions than those
previously  proposed in \cite{Donoho2006Stable},
\cite{Cai2011Orthogonal} as soon as $\x_\Qcs$ is not a flat vector. 
   For flat vectors, \eqref{eq:condn4} obviously
reduces to \eqref{eq:condnoisystd2}. 
 In such a case,  Theorem  \ref{corr:SRCnoisy} leads to the standard conditions by Donoho \etal

Let us mention that the conditions in Theorem \ref{corr:SRCnoisy} do
not enforce any constraint on the decay of the coefficients in $\Qcs$
(but only between the elements in $\Qcsc$ and each component of
$\Qcs$). In the next theorem, we state ``horizon-1" conditions of the
same flavor as those presented in Theorems \ref{th:mainresult2} and
\ref{th:mainresult3}.  
 Let us first define the following quantity:
%
\begin{align}\label{eq:defgamma}
\gamma_{k}\triangleq\left\{
\begin{array}{ll}
\frac{1-(k-2)\mu}{(\mu+1)(1- k\mu)} & \mbox{for OMP},\\
\sqrt{
\frac{1-(k-2)\mu}{\mu+1}} \frac{\sqrt{1- (k-1)\mu}}{1- k \mu} & \mbox{for OLS}.
\end{array}
\right. 
\end{align}
Our result then writes as follows: 
\begin{theorem}\label{th:noisymainresult}
If
\begin{align}
\mu < \frac{1}{k}, \label{eq:condn1}
\end{align}
and 
 $\forall i\in\{1,\ldots,k\}$,
\begin{align} \label{eq:condn2}
\vert x_i \vert &> \frac{2\mu (k-i)}{1-i \mu} \vert x_{i+1}\vert + 2
\gamma_{k}(\epsilon+\|\x_\Qcsc\|_1),
\end{align}
then Oxx selects atoms in $\Qcs$ from noisy data during the first $k$ iterations. 
\end{theorem}

Theorem \ref{th:mainresult3} could be extended in a similar way to the
framework of compressible and noisy signals but we do not detail this
extension for conciseness.


 \section{Technical Details}\label{sec:td}

In this section, we provide a proof of the theorems stated in section \ref{sec:Results}. We first recall the main principles ruling OMP and OLS in section \ref{sec:Oxx}. We then introduce some technical lemmas in section \ref{sec:usefullemmas}. Finally the proof of the main results is exposed in section \ref{sec:proofmth}.  


\subsection{OMP and OLS} \label{sec:Oxx}


In order to precisely describe the update rules characterizing Oxx,
let us first introduce some notations: given a set of indices $\Qc$,
$\A_\Qc$ represents the submatrix of $\A$ specified by the columns
indexed in $\Qc$; the projector onto the orthogonal complement of the
column span of $\A_\Qc$ is defined as $\proj\triangleq\I-\A_\Qc \A_\Qc^\dag$, where $\A_\Qc^\dag$ is the pseudo-inverse of $\A_\Qc$; in particular, $ \r^\Qc \triangleq \proj \y$ is the residual error when projecting $\y$ onto the span of $\A_\Qc$.  Finally, $\langle \cdot, \cdot  \rangle$ represents the vector inner product and
$\mathbf{0}_m$ is the null vector of size $m\times 1$. 

Oxx can be understood as an iterative procedure generating an estimate of $\Qcs$ by sequentially adding one new element to the current support estimate, say $\Qc$. As detailed in Algorithm~\ref{alg:Oxx}, OMP and OLS differ in the way this new element is selected. 
 At each iteration, OLS selects the atom $\a_j$ yielding the minimum
residual error $\|\r^{\Qc\cup\{j\}}\|_{2}$:
\begin{equation*}
  j \in\argmin_{i\notin\Qc}\|\r^{\Qc\cup\{i\}}\|_{2},
  \label{eq:ols_rule_ai}
\end{equation*}
and $n-\mathrm{Card}\{\Qc\}$ least-square problems have to be solved to compute
$\|\r^{\Qc\cup\{i\}}\|_{2}$ for all $i\notin\Qc$~\cite{Chen:1950fk}. On the
contrary, OMP adopts the simpler rule
\begin{equation*}
j \in\argmax_{i\notin\Qc}|\langle \a_i, \r^\Qc \rangle |,
  \label{eq:omp_rule_ai}
\end{equation*}
to select the new atom $\a_j$, and then solves only one
least-square problem to update the new residual 
$\r^{\Qc\cup\{j\}}$.

The selection rules described above can also be expressed in terms of the (normalized) projected atoms of the dictionary \cite{Soussen2013Joint}. This formulation will turn out to be convenient in our proofs below. More specifically, let 
\begin{align}
\ta_i & \triangleq \P_{\Qc}^\perp \a_i, \nonumber\\
\tb_i & \triangleq \left\{
\begin{array}{ll}
{\ta_i}/{\| \ta_i \|_{2}}& \mbox{if $\ta_i\neq\mathbf{0}_m$}\\
\mathbf{0}_m & \mbox{otherwise.}
\end{array}\right. \nonumber
\end{align}
With these notations, the selection rule of Oxx can be re-expressed as (see \eg \cite{RebolloNeira2002Optimized})
\begin{align}\label{eq:atomselection2}
 j \in \argmax_{i\notin\Qc}  \vert \langle \tc_i,\r^{\Qc} \rangle \vert,
\end{align} 
where 
\begin{align}\nonumber
\tc_i \triangleq
  \left\{
    \begin{array}{ll}
      \ta_i & \textrm{for OMP},\\
      \tb_i & \textrm{for OLS}.
    \end{array}
  \right.
\end{align}
 For simplicity, the dependence of $\ta_i$, $\tb_i$ and $\tc_i$ on $\Qc$ does not appear in our notations. The reader should however keep this dependence in mind in our subsequent derivations.


\subsection{Some Useful Lemmas} \label{sec:usefullemmas}

We first state three useful lemmas, connecting different functions of the projected atoms to the mutual coherence of the dictionary.  
\begin{lemma}\label{lemma:laumu}
Let $\mathrm{Card}(\Qc)=g\geq 0$. If $\mu<\frac{1}{g}$, then
\begin{align}
\begin{array}{ll}
\|\ta_i \|^2_{2}\geq \frac{(\mu+1)\,(1-g\mu)}{1-(g-1)\mu}&\quad \forall i\notin \Qc,\\
\vert\langle  \ta_i, \ta_j \rangle\vert \leq \frac{\mu\,(\mu+1)}{1-(g-1)\mu}&\quad \forall j\neq i.
\end{array}
\label{eq:lem1}
\end{align}
\end{lemma} 

\emph{Proof:}
  The result is a direct consequence of Lemmas 4 and 10 in \cite{Herzet2013}.
\hfill$\square$\\

\begin{lemma}\label{lemma:alphabeta}
Let $\mathrm{Card}(\Qc)=g\geq 0$.  If $\mu<\frac{1}{g}$, we have
%
\begin{align}\nonumber
\begin{array}{ll}
 \langle  \tc_i, \ta_i  \rangle\geq \alpha_g>0 &\quad \forall i\notin\Qc,\\
\vert\langle  \tc_i, \ta_j \rangle\vert \leq \mu_g & \quad  \forall j\neq i,
\end{array}
\end{align}
where
%
%
\begin{align}
\alpha_g =& \label{eq:alphadef}
\left\{
\begin{array}{ll}
 \frac{(\mu+1)\,(1-g\mu)}{1-(g-1)\mu} & \mbox{for OMP} \\
\sqrt{ \frac{(\mu+1)(1-g\mu)}{1-(g-1)\mu}} & \mbox{for OLS}
\end{array}
\right.\\
\mu_g=& \label{eq:mudef}
\; \min \left \{1,\,\frac{\mu}{1-g\mu}\;\alpha_g \right
\}.
\end{align}
%
%
\end{lemma}
\emph{Proof:} 
The result immediately follows from Lemma \ref{lemma:laumu}  and
  from $\|\ta_j\|_2\leq \|\a_j\|_2=1$ and $\|\tc_i\|_2\leq 1$. Note that
$\mu<\frac{1}{g}$ implies that $\ta_i\ne \mathbf{0}_m$
(see~\eqref{eq:lem1}).  Thus, $\tb_i$ reads ${\ta_i}/{\| \ta_i
  \|_{2}}$.
\hfill$\square$\\

\begin{lemma}\label{lemma:alphabetaa}
If $\mu\leq\frac{1}{g+1}$, then
\eqref{eq:mudef} simplifies to: 
\begin{align}
\mu_g=
\frac{\mu}{1-g\mu}\;\alpha_g.
\label{eq:mudef_simpl}
\end{align}
\end{lemma}

\emph{Proof:} 
 $\mu\leq\frac{1}{g+1}$ implies that $\mu\leq 1-g\mu$ and then 
\begin{align*}
%
\frac{\mu}{1-g\mu}\alpha_g\leq \alpha_g\leq 1,
\end{align*}
where  $\alpha_g\leq 1$ follows from  $\forall i\notin\Qc$, 
$\alpha_g\leq  \langle  \tc_i, \ta_i  \rangle \leq 
\|\tc_i\|_2\|\ta_i\|_2\leq 1$.
\hfill$\square$\\

Lemmas \ref{lemma:alphabeta} and \ref{lemma:alphabetaa} are the building blocks of the next lemma, which provides sufficient conditions for Oxx to select a good atom at a given iteration: 

\begin{lemma}\label{lem:mainlemma}
Consider a (possibly non-sparse) signal $\x$ and a subset $\Qcs$ of cardinality $k$. 
Assume that Oxx, with $\y$ defined as in \eqref{eq:defy} as input, has selected atoms in $\Qc\subsetneq \Qcs$ during the first $g$ iterations, with $0\leq g <k$.
Let
 $\alpha_g$, $\mu_g$ be defined as in Lemma \ref{lemma:alphabeta}.
 If 
%
%
%
\begin{align}
&\mu< \frac{1}{g},\label{eq:mainlemma0}\\
%
&(\alpha_g + \mu_g) \| \x_{\Qcs\backslash\Qc}\|_\infty - 2 \mu_g \| \x_{ \Qcs\backslash\Qc}\|_1 > 2\, (\epsilon + \|\x_\Qcsc\|_1), \label{eq:mainlemma} 
\end{align}
then Oxx selects an atom in $\Qcs\backslash\Qc$ at the next iteration. 
\end{lemma}

\vspace{0.2cm}
%
\emph{Proof :} We want to show that \eqref{eq:mainlemma0}-\eqref{eq:mainlemma} implies
\begin{align}
\displaystyle{\max_{i\in\Qcs\backslash\Qc}} \vert\langle \tc_i, \r^\Qc\rangle \vert > \vert\langle \tc_l, \r^\Qc\rangle\vert, \quad \forall l\notin \Qcs. \label{proof:mgoal}
\end{align}
First, using the definitions of the residual $\r^\Qc=\P_{\Qc}^\perp \y$ and the projected atoms $\ta_i$, we have
\begin{align}\nonumber
\r^\Qc = \s^\Qc + \proj \w,
\end{align}
where
\begin{align}\nonumber
\s^\Qc 
= \sum_{i\notin \Qc} \ta_i \, x_i.
\end{align}
Noticing that  $\|\tc_i\|_2\leq 1$ and $\|\P_{\Qc}^\perp \w\|_2\leq\|\w\|_2\leq\epsilon$, a sufficient condition for \eqref{proof:mgoal} is then as follows:
\begin{align}
\displaystyle{\max_{i\in\Qcs\backslash\Qc}} \vert\langle \tc_i, \s^\Qc\rangle \vert - \vert\langle \tc_l, \s^\Qc\rangle\vert> 2\epsilon,\quad \forall l\notin \Qcs. \label{proof:mgoal2}
\end{align}
 Let $j\in\argmax_{i\in \Qcs\backslash\Qc} \vert x_{i} \vert$. Since $\mu<\frac{1}{g}$, we can apply Lemma \ref{lemma:alphabeta}  and bound the terms in the left-hand side of \eqref{proof:mgoal2} as follows: 
\begin{align}
\displaystyle{\max_{i\in\Qcs\backslash\Qc}} \vert\langle \tc_i, \s^\Qc\rangle \vert
 &\geq \vert\langle \tc_j, \s^\Qc\rangle \vert\nonumber\\
 &\geq  \vert\langle \tc_j, \ta_j\rangle \vert\, |x_j| - \sum_{i\notin\Qc\cup\{j\}} \vert\langle \tc_j, \ta_i \rangle \vert |x_i |\nonumber\\
 &\geq  \alpha_g\, |x_j| - \mu_g\, (\| \x_{\Qcs\backslash(\Qc\cup\{j\})} \|_1+\|\x_\Qcsc\|_1), \nonumber
\end{align}
and $\forall l\notin\Qcs$, 
\begin{align}
 \vert\langle \tc_l, \s^\Qc\rangle\vert
 &\leq \vert\langle \tc_l, \ta_l\rangle\vert\, \vert x_l\vert + \sum_{i \notin \Qc\cup\{l\}} \vert\langle \tc_l, \ta_i \rangle \vert |x_i|\nonumber\\
 &\leq \vert x_l\vert + \mu_g\, (\| \x_{\Qcs\backslash\Qc}\|_1+\| \x_{\Qcsc\backslash\{l\}}\|_1),\nonumber
\end{align}
where the last inequality follows from the fact that $\vert\langle \tc_l, \ta_l\rangle\vert\leq 1$. 
 Combining these two bounds, we easily obtain that 
 \begin{align}
  (\alpha_g + \mu_g) |x_j | - 2 \mu_g& \| \x_{ \Qcs\backslash\Qc }\|_1 \nonumber\\
  &> 2\, \epsilon + (1-\mu_g) |x_l | +2 \mu_g \|\x_{\Qcsc}\|_1\nonumber
\end{align}
 is a sufficient condition for
\eqref{proof:mgoal2} and then 
 \eqref{proof:mgoal}. Finally, noticing that
  $\mu_g\leq 1$ (Lemma 2)
 and $|x_l |\leq \|\x_{\Qcsc}\|_1$, 
  we obtain \eqref{eq:mainlemma}.
  \hfill$\square$\\

\subsection{Proofs of the Main Results} \label{sec:proofmth}

In this section, we provide a proof of the main theorems of the
paper. We skip the proofs of the corollaries, which are
straightforward.  Theorems \ref{eq:relaxation_theorem},
\ref{th:mainresult3}, \ref{corr:SRCnoisy}, \ref{th:noisymainresult}
and the direct part of Theorem \ref{th:mainresult2} are proved in
section \ref{sec:proofdirectparts}. The converse part of Theorem
\ref{th:mainresult2} (that is the tightness of the proposed
conditions) is proved in section
\ref{sec:proofnth}.\\
%
%

\subsubsection{Proofs of the Sufficient Conditions}\label{sec:proofdirectparts}

All the proofs of this part use Lemma \ref{lem:mainlemma} as a key building block.\\

\emph{Proof of Theorem \ref{eq:relaxation_theorem}: } We want to show that Oxx selects atoms in $\Qcs$ during the  first $k$ iterations for all  dictionaries obeying $\mu<\mu^\star$, for some $\mu^\star>\frac{1}{2k-1}$,  as long as $\x_\Qcs$ is not a flat vector. 

Let us first derive a condition on the mutual coherence ensuring that Oxx makes a correct decision at the \textit{first} iteration. Particularizing the sufficient conditions of Lemma \ref{lem:mainlemma}  to the case $\Qc=\emptyset$ (with $\epsilon=0$ and $\|\x_\Qcsc\|_1=0$), we have that Oxx selects an element of $\Qcs$ provided that:
\begin{align}\nonumber
\mu<\frac{\rho}{2 -\rho},
\end{align}
with 
$\rho=\| \x_{\Qcs} \|_\infty/\|\x_\Qcs\|_1$. 
%
%
Now, since $\x_\Qcs$ is not flat, we have that $\rho>\frac{1}{k}$,
%
%
and therefore 
$\frac{\rho}{2 - \rho}>\frac{1}{2k-1}$.

On the other hand, if Oxx has selected any $g\geq 1$ atoms  in $\Qcs$
during the first $g$ iterations, it was proved in
\cite[Th. 3]{Herzet2013} that Oxx makes good decisions during the
remaining $k-g$ iterations provided that
$\mu<\frac{1}{2k-g-1}$.
A sufficient condition for Oxx to select correct atoms during the first $k$ steps thus simply writes
\begin{align}\nonumber
\mu<\mu^\star
\ \mbox{ with }\ 
\mu^\star &= \min\left(\frac{\rho}{2 -\rho},\frac{1}{2k-2}\right).\nonumber
\end{align}
Clearly, $\mu^\star>\frac{1}{2k-1}$ by definition. 
\hfill$\square$\\[0.2cm]

The direct part of Theorem \ref{th:mainresult2} can be seen as a
special case of Theorem \ref{th:noisymainresult} when $\epsilon=0$ and
 $\x_\Qcsc=\mathbf{0}_{n-k}$,
  and of Theorem~\ref{th:mainresult3} with $g=0,r=p=k$. Hence, we focus
  on the latter proofs hereafter.  \\

\emph{Proof of Theorem \ref{th:noisymainresult}:}  Assume that Oxx has selected atoms in $\Qc\subsetneq\Qcs$ when $g\leq k-1$ iterations have been completed; we apply Lemma \ref{lem:mainlemma} to show that, under the hypotheses of Theorem~\ref{th:noisymainresult}, the next atom selected by Oxx belongs to $\Qcs\backslash\Qc$.

%

The first condition of Lemma \ref{lem:mainlemma}, $\mu<\frac{1}{g}$, is always verified since $\mu<\frac{1}{k}$ by hypothesis and  $g\leq k-1$.
Let $j$ be the lowest index such that: 
\begin{align}
j&\in\argmax_{i\in\Qcs\backslash\Qc} |x_i|.\label{eq:sideeqth1b}
\end{align}
 Clearly, \eqref{eq:h2} implies that $j\leq g+1$. 

Because the nonzero
coefficients have been sorted in the decreasing order, see
 \eqref{eq:h2}, we have: 
\begin{align}
\| \x_{\Qcs\backslash\Qc}\|_1
& \leq |x_j|+(k-g-1)\, |x_{j+1}|. \label{eq:boundl1n}
\end{align}
Hence, 
\begin{align}
(\alpha_g-\mu_g) \vert x_{j} \vert&- 2 \mu_g (k-g-1) |x_{j+1}| > 2\,(\epsilon+\|\x_{\Qcsc}\|_1) \label{eq:popoooopo}
\end{align}
is a sufficient condition for \eqref{eq:mainlemma}. Since 
 $g\leq k-1$
 and $\mu<\frac{1}{k}$ by assumption, we have
$\mu<\frac{1}{g+1}$ and we can exploit the expression 
 \eqref{eq:mudef_simpl} of $\mu_g$ in Lemma~\ref{lemma:alphabetaa} to rewrite:
\begin{align}
\label{eq:34}
\alpha_g-\mu_g&=\alpha_g\,\left (1-\frac{\mu}{1-g\mu} \right)>0.  
\end{align}
%
It follows from~\eqref{eq:34} that
\begin{align*}
\frac{\mu_g}{\alpha_g-\mu_g}&=\frac{\alpha_g}{\alpha_g-\mu_g}-1,\\
&=\frac{1-g\mu}{1-(g+1)\mu}-1,\\
&=\frac{\mu}{1-(g+1)\mu}.
\end{align*}
Then, \eqref{eq:popoooopo} can be rewritten as
\begin{align}
\vert x_j \vert > \frac{2\mu (k-g-1)}{1-(g+1) \mu} \vert x_{j+1}\vert
+
\frac{2(\epsilon+\|\x_{\Qcsc}\|_1)}{\alpha_g-\mu_g}.
\end{align}
We finally obtain
condition \eqref{eq:condn2} 
by noticing that:
\begin{itemize}
  \item $j\leq g+1$
and the function $f(u)=\frac{2\mu (k-u)}{1-u \mu}$
 is decreasing on
$u\in[0,k]$
 for $\mu<\frac{1}{k}$; 
 \item $\alpha_g$ (see~\eqref{eq:alphadef}) and
     $1-\frac{\mu}{1-g\mu}$ are both decreasing with $g$ and
     non-negative, hence $1/(\alpha_g-\mu_g)$ (see~\eqref{eq:34})
     is increasing with $g$. It is upper bounded by
     $1/(\alpha_{k-1}-\mu_{k-1})$, which is equal to $\gamma_{k}$
     defined in~\eqref{eq:defgamma}.\\
\end{itemize}
We can thus conclude that \eqref{eq:condn1}-\eqref{eq:condn2} are sufficient conditions for  \eqref{eq:mainlemma0}-\eqref{eq:mainlemma} and the next atom selected by Oxx belongs to $\Qcs\backslash\Qc$ by virtue of Lemma \ref{lem:mainlemma}. 

This proof applies recursively to the iterations of Oxx for increasing
values of $g\in\{0,\ldots,k-1\}$. 
\hfill$\square$\\[0.2cm]

The proofs of Theorems \ref{th:mainresult3} and \ref{corr:SRCnoisy} follow a reasoning in the same vein as Theorem \ref{th:noisymainresult} but with some variations that we describe below. \\

\emph{Proof of Theorem \ref{th:mainresult3} (First Part):} By hypothesis, we assume
that Oxx has selected atoms in  $\Qc\subsetneq\Qcs$ during the
first $g$ iterations. We recursively show that, if \eqref{th:th3.a} and \eqref{th:th3.b} are satisfied, then Oxx keeps on picking atoms
in $\Qcs$ as long as the $p$ largest
elements of $\x_{\Qcs \backslash\Qc}$ have not been
selected.

Assume that after iteration
$g+t$, $t\in\{0,\ldots,k-g-1
  \}$, has been completed, Oxx has selected $\Qc'$ with
$\Qc\subseteq\Qc'\subsetneq\Qcs$ and
%
 $\{1, \ldots, p\} \nsubseteq \Qc'$
 (that is, some atoms in $\{1, \ldots, p\}$ have not yet been
   selected by Oxx). We apply Lemma
 \ref{lem:mainlemma} (with subset $\Qc\leftarrow \Qc'$, and
  with $\epsilon\leftarrow 0$,
$\x_\Qcsc\leftarrow\mathbf{0}_{n-k}$) to
prove that the next atom selected by Oxx belongs to $\Qcs$.
The first condition in Lemma \ref{lem:mainlemma},
$\mu<\frac{1}{g+t}$, is verified since $\mu<\frac{1}{k}$ and
 $g+t< k$.
 
Let $j$ be the lowest index such that 
\begin{align}
j&\in\argmax_{i\in\Qcs\backslash\Qc'} |x_i|.\label{eq:sideeqth1b2}
\end{align}
%
%
 Because of our assumption $\{1, \ldots, p\} \nsubseteq \Qc'$
  we necessarily have that
  $\{1, \ldots, p\}\cap (\Qcs\backslash\Qc')\ne\emptyset$, and then
  from convention \eqref{eq:h2B}, we must have $j\in\{1,\ldots,p\}$.
By the same arguments as those exposed in the proof of Theorem \ref{th:noisymainresult}, we have that
%
%
\begin{align}\label{eq:tmp}
\vert x_j \vert > \frac{2\mu (k-g-t-1)}{1-(g+t+1) \mu} \vert x_{j+1}\vert
\end{align}
is a sufficient condition for \eqref{eq:mainlemma}.  Next, we have
 from the convention~\eqref{eq:h2B}
that $j\leq t+1$, and the function 
  $f(u)=\frac{2\mu (k-u)}{1-u \mu}$
 is decreasing on
$u\in[0,k]$, hence on $u\in[g+j,g+t+1]\subseteq[0,k]$ for $\mu<1/k$.
So, 
\begin{align}
\vert x_j \vert > \frac{2\mu (k-g-j)}{1-(g+j) \mu} \vert x_{j+1}\vert\label{eq:condpartialrecovery}
\end{align}
is sufficient for \eqref{eq:tmp}  and then \eqref{eq:mainlemma}. 
%
%
%
Since $j\in\{1,\ldots,p\}$, \eqref{eq:condpartialrecovery}  holds by
virtue of \eqref{th:th3.b} and the next atom selected by Oxx belongs to $\Qcs$.  \hfill$\square$\\[0.2cm]

\emph{Proof of Theorem \ref{th:mainresult3} (Second Part):} We show
that \eqref{th:th3.c} and \eqref{th:th3.d}  ensure that
 atoms in $\Qcs\backslash \Qc$ are selected, provided that the $p$ largest
elements of $\x_{\Qcs \backslash\Qc}$ have not been
selected and less than $g+r$ iterations have been carried out. The proof follows the same lines as the proof of the first part with some modifications that we describe hereafter. 
 
  Assume that Oxx has selected $\Qc'$ with
$\Qc\subseteq\Qc'\subsetneq\Qcs$ after iteration
$g+t$ 
has been completed, with $t\in\{0,\ldots,r-1
  \}$, and $\{1, \ldots, p\} \nsubseteq \Qc'$.  We apply
Lemma \ref{lem:mainlemma} (with subset $\Qc\leftarrow \Qc'$,
 $\epsilon\leftarrow 0$ and
$\x_\Qcsc\leftarrow\mathbf{0}_{n-k}$) to
show that the next atom selected by Oxx belongs to $\Qcs$. 
 We note that the first condition of Lemma \ref{lem:mainlemma},
 $\mu<\frac{1}{g+t}$, is satisfied since $\mu<\frac{1}{g+r}$ by
hypothesis and $t<r$. 

Denoting by $j$ be the lowest index such that \eqref{eq:sideeqth1b2} is verified, we have by virtue of Lemma \ref{lem:mainlemma} that \eqref{eq:tmp} is a sufficient condition for Oxx to select an atom in $\Qcs$ at the next iteration. From the same arguments as in proof of the first part, we must have $j\in\{1,\ldots, p\}$. 

Now, if $\frac{1}{k}\leq \mu < \frac{1}{g+r}$, the function 
  $f(u)=\frac{2\mu (k-u)}{1-u \mu}$ is nondecreasing on 
 $u\in [g+t+1, g+r]\subseteq [0,g+r]$. As a consequence,
\begin{align}
\vert x_j \vert > \frac{2\mu (k-g-r)}{1-(g+r) \mu} \vert x_{j+1}\vert\label{eq:condpartialrecovery2}
\end{align}
 is a sufficient condition for \eqref{eq:tmp}  and then \eqref{eq:mainlemma}.

%

Since $j\in\{1,\ldots,p\}$, \eqref{eq:condpartialrecovery2} holds by
virtue of \eqref{th:th3.d} and the next atom selected by Oxx belongs to $\Qcs$.
\hfill$\square$\\

\textit{Proof of Theorem \ref{corr:SRCnoisy}:} We first note that
$\mu<\frac{1}{g+1}$ $\forall g\in \{0,\ldots, k-1\}$ since
$\mu<\frac{1}{2k-1}$ by hypothesis and $2k-1\geq k\geq g+1$. The
beginning of the proof follows the same lines as the proof of
Theorem \ref{th:noisymainresult} 
 but exploits that, by
assumption  
 \eqref{eq:h2},
 $|x_{j}| \geq |x_{j+1}|$ for $j<k$. 
Let $j\leq g+1$ be defined in \eqref{eq:sideeqth1b}.
 If $j<k$, \eqref{eq:h2} implies that:
\begin{align}
(\alpha_g-\mu_g(2k-2g-1))  |x_{j}| > 2\, (\epsilon+\|\x_\Qcsc\|_1)
\label{eq:neweq}
\end{align}
is a sufficient condition for \eqref{eq:popoooopo}. 
 The same result holds in the case $j=k$, since then, $g=k-1$
and~\eqref{eq:neweq} identifies with \eqref{eq:popoooopo}.

 Since $\mu<\frac{1}{g+1}$, Lemma \ref{lemma:alphabetaa} applies and we can use 
\eqref{eq:mudef_simpl} to rewrite
the latter condition as
\begin{align}
\frac{\alpha_g}{1-g \mu}
(1-(2k-g-1)\mu)  |x_{j}| > 2\, (\epsilon+\|\x_\Qcsc\|_1).
\label{eq:intcondth3}
\end{align}
It can be easily checked from the definition of $\alpha_g$ in
\eqref{eq:alphadef} that $\alpha_g\geq 1-g\mu$ holds whenever
$\mu<1/g$.  As a consequence, \eqref{eq:intcondth3} can be relaxed as
%
%
\begin{align}
\left(1-(2k-g-1)\mu \right) |x_j|>2(\epsilon+\|\x_\Qcsc\|_1). \nonumber
\end{align}
Finally, since $\mu<\frac{1}{2k-1}\leq \frac{1}{2k-g-1}$ and
  $|x_j|\geq |x_{g+1}|$, this condition can be 
relaxed as
\eqref{eq:condn4}. Hence, the atom selected by Oxx at iteration $g+1$
belongs to $\Qcs\backslash\Qc$ by virtue of Lemma \ref{lem:mainlemma}.
\hfill$\square$\\

\subsubsection{Proof of the converse part of Theorem \ref{th:mainresult2}} \label{sec:proofnth}
 In this section, we prove that the conditions defined in
 \eqref{eq:cond1}-\eqref{eq:cond2}  are worst-case necessary for
 the success of Oxx in the  sense specified in Theorem
 \ref{th:mainresult2}.  Our proof is based on the construction of 
 an equiangular  dictionary
$\A\in \mathbb{R}^{(k+1)\times (k+1)}$ and a $k$-sparse vector
$\x\in\R^{k+1}$ leading to a failure of Oxx during the first $k$ steps. 
 More specifically, we show that the following type of situation  occurs:
\begin{align}\label{eq:OxxAmbiguityFailure}
\max_{i\in \Qcs} \vert \langle \tc_i , \r^\Qc \rangle\vert = \vert \langle \tc_j , \r^\Qc \rangle\vert \quad \mbox{\ for some $j\notin \Qcs$,}
\end{align}
that is, there is  an ambiguity in the choice of the next atom. In such a case, Oxx cannot be ensured to select a  good atom.  

Let $\G \in\mathbb{R}^{(k+1) \times (k+1)}$ be a matrix with ones on the diagonal and 
$-\mu$, with $\mu\leq\frac{1}{k}$, elsewhere. $\G$ will play the
  role of the Gram matrix $\G=\A^T\A$. We will exploit the eigenvalue
decomposition of $\G$ to construct a dictionary
$\A\in\mathbb{R}^{(k+1)\times (k+1)}$ with the desired properties.
Since $\G$ is symmetric, it can be expressed as
\begin{align}\nonumber
\G = \U \Diag \U^T,
\end{align}
where $\U$ (respectively, $\Diag$) is the unitary matrix whose
  columns are the eigenvectors (respectively, the diagonal matrix of
eigenvalues) of $\G$. It is easy to check that $\G$ has only two
distinct eigenvalues: 
$1+\mu$ with multiplicity $k$ and $1-k\mu$ with multiplicity one.

Because we assume $\mu\leq\frac{1}{k}$, all the diagonal elements of $\Lambda$ are non-negative and the latter matrix can be factorized as $\Lambda=\Lambda^{1/2} \Lambda^{1/2}$. The dictionary $\A$ can then be defined as 
\begin{align}\label{eq:dicodef}
\A=\Lambda^{1/2}\U^T. 
\end{align}
We note that by definition, we have 
 $\A^T \A = \G$, 
and therefore 
\begin{align}
\vert \langle \a_i, \a_j \rangle\vert = \mu \quad \forall i \neq j.
\label{eq:prod_scal_atomes}
\end{align}
With this choice of dictionary, the following lemma holds:
\begin{lemma}\label{lem:equiv_alphag_mu}
Consider the dictionary $\A$ described above with
$\mu\leq\frac{1}{k}$. Then, we have for all $\Qc$ with $\mathrm{Card}(\Qc)=g\leq k-1$:
\begin{align}\label{eq:alphacst}
\begin{array}{ll}
 \langle \tc_i,\ta_i\rangle = \alpha_g&\quad \forall\,i \notin\Qc,\\
 \langle \tc_i,\ta_j\rangle = -\mu_g&\quad \forall\,i,j\notin\Qc,i\neq j,
\end{array}
\end{align}
where $\alpha_g$ and $\mu_g$ are defined in \eqref{eq:alphadef}-\eqref{eq:mudef_simpl}. 
\end{lemma}

\textit{Proof:}  The expression \eqref{eq:mudef_simpl} of $\mu_g$ holds since 
$\mu\leq 1/(g+1)$.

The result \eqref{eq:alphacst} is obvious for $\Qc=\emptyset$ since $\alpha_0=1,
\mu_0=\mu$ and  $\forall\,i$, $\tc_i=\ta_i=\a_i$. We now address the
case where $\Qc\ne\emptyset$.

First notice that because $\mathrm{Card}(\Qc)=g\leq k-1$, the
matrix $\A_\Qc^T\A_\Qc$ is invertible. Indeed, it is a symmetric matrix with ones on the diagonal
and $-\mu$ elsewhere. Its eigenvalues (\ie $1+\mu$ with multiplicity
$g-1$ and $1-(g-1)\mu$ with multiplicity one) are therefore strictly
positive if $\mu\leq\frac{1}{k}$. Hence, the  projected atoms $\ta_i^{}$ can be expressed as
\begin{align}\nonumber
\ta_i^{} 
&= \a_i - \A_{\Qc} (\A_{\Qc}^T \A_{\Qc})^{-1} \A_{\Qc}^T \a_i.
\end{align}
Using this expression, we also have
\begin{align}\nonumber
\stdscal{\ta_i^{},\ta_j^{}}&=\stdscal{\a_i,\a_j}-\a_i^T\A_{\Qc} (\A_{\Qc}^T \A_{\Qc})^{-1} \A_{\Qc}^T \a_j,\\
\| \ta_i^{}\|^2_2&= 1- \a_i^T\A_{\Qc} (\A_{\Qc}^T \A_{\Qc})^{-1} \A_{\Qc}^T \a_i.\nonumber
\end{align}
Taking into account that the inner product between any pair of 
distinct atoms is equal to $-\mu$ by definition of
$\G=\A^T\A$, we obtain for $i,j\notin\Qc$: 
\begin{align} \label{eq:eqintprooflemme5}
\begin{array}{l}
\stdscal{\ta_i^{},\ta_j^{}}=-\mu-\mu^2\mathbf{1}_{g}^T (\A_\Qc^T \A_\Qc)^{-1} \mathbf{1}_{g},\\
\| \ta_i^{}\|^2_2= 1-\mu^2\mathbf{1}^T_{g} (\A_\Qc^T \A_\Qc)^{-1} \mathbf{1}_{g},
\end{array}
\end{align}
where $\mathbf{1}_{g}$ denotes the ``all-ones" vector of dimension
$g$. 

Finally, we obtain the result for OMP ($\tc_i=\ta_i$) by
noticing 
that $\onebf_{g}$ is an eigenvector of $(\A_\Qc^T \A_\Qc)^{-1}$
with eigenvalue $\frac{1}{1-(g-1)\mu}$ and identifying the right-hand sides in \eqref{eq:eqintprooflemme5} with \eqref{eq:alphadef} and~\eqref{eq:mudef_simpl}. The proof
  of~\eqref{eq:alphacst} for OLS ($\tc_i=\tb_i$) is a direct consequence
of~\eqref{eq:alphacst} in the case of OMP, using $\tb_i = \ta_i/\|\ta_i\|_2$.
  \hfill$\square$\\

To prove the tightness of each of the conditions 
\eqref{eq:cond2}, 
we need to introduce  particular instances $\x^{(j)}$ of $\x\in\mathbb{R}^{k+1}$, for $j=1,\ldots,k-1$:
%
\begin{align}\label{eq:thevampireslayer}
 &\left\lbrace
\begin{array}{lll}
x_i^{(j)}   &>  \frac{2\mu (k-i)}{1-i \mu}  x_{i+1}^{(j)} & \mbox{for $1\leq  i < j $},\\
 x_j^{(j)}  &=  \frac{2\mu (k-j)}{1-j \mu}  x_{j+1}^{(j)} &\\
 x_i^{(j)}  &=  1&\mbox{for  $j+1 \leq i \leq k$},\\
x_{k+1}^{(j)} &= 0. & 
\end{array}
\right.
\end{align}

%
\begin{lemma}\label{lem:goodatomselecCN}
Consider the dictionary $\A$ described above with
$\mu =\frac{1}{2k-j}$ and 
$\x^{(j)}\in\mathbb{R}^{k+1}$ defined as in
\eqref{eq:thevampireslayer}, for $j\leq k-1$. Then, the ordering~\eqref{eq:h2} holds. Moreover, for $g\in\{0,\ldots,j-1\}$, let
$\Qc=\{1,\ldots,g\}$. Then, $\langle
\tc_{g+1},\r^{\Qc} \rangle\geq 0$ and
%
%
\begin{align} \label{eq:lemma5main}
(g+1)\in\argmax_{i\notin{\Qc}} 
|\langle \tc_i,\r^{\Qc} \rangle|.
\end{align}
If $g<j-1$, then $g+1$ is the unique minimizer of \eqref{eq:lemma5main}. 
\end{lemma}

\textit{Proof}: 
Let us first notice that $\mu=\frac{1}{2k-j}$ ensures that our working
assumption \eqref{eq:h2} is met because $\frac{2\mu (k-i)}{1-i
  \mu}\geq1$ for $i \leq j$.  Note also that $\x^{(j)}$ is
non-negative.

 Since $\mu=\frac{1}{2k-j}$ and $j\leq k-1$, we have $\mu\leq \frac{1}{k+1}<\frac{1}{g+1}$ and  Lemma~\ref{lemma:alphabetaa} applies, leading to $\frac{2\mu_{g}}{\alpha_{g}- \mu_{g}}=
\frac{2\mu}{1-(g+1)\mu}>0$.
By virtue of \eqref{eq:thevampireslayer}, we have:
\begin{align}\label{eq:keyfeatx}
(\alpha_{g}-\mu_{g}) x_{g+1}^{(j)} -2
\mu_{g} (k-g-1) x_{g+2}^{(j)}
\left\{
\begin{array}{l}
>0\;\;\textrm{for}\; g<j-1,\\
=0\;\;\textrm{for}\; g=j-1.
\end{array}
\right .
\end{align}
%

We first prove that $\langle \tc_{g+1},\r^{\Qc} \rangle\geq
0$, 
$\forall g\leq j-1$. Indeed, applying Lemma \ref{lem:equiv_alphag_mu}, we have
\begin{align}
\langle \tc_{g+1},\r^{\Qc} \rangle 
&= \langle \tc_{g+1},\sum_{l\notin\Qc} x_l^{(j)}\ta_l \rangle \nonumber\\
&= \alpha_{g} x_{g+1}^{(j)} - \mu_{g} \sum_{l\notin\Qc\cup\{g+1\}} x_l^{(j)}\nonumber\\
&\geq \alpha_{g} x_{g+1}^{(j)} - \mu_{g} (k-g-1) x_{g+2}^{(j)} \nonumber\\
&\geq 0, \label{eq:ineqg+1}
\end{align}
where the first inequality follows from \eqref{eq:h2} and the second
from \eqref{eq:keyfeatx}
 and $x_i^{(j)}\geq 0$ for all $i$.

We then show that \eqref{eq:lemma5main} holds by proving that :
\begin{align}\label{eq:theselemma5}
\langle \tc_{g+1},\r^{\Qc} \rangle-|\langle \tc_{i},\r^{\Qc} \rangle| \geq 0,
\;\; \forall i \notin\Qc\cup\{g+1\}.
\end{align}
%
For $s\in\{-1,1\}$, let $f_i(s)\triangleq \langle \tc_{g+1},\r^{\Qc}
\rangle - s\langle \tc_{i},\r^{\Qc} \rangle $.  
Obviously, \eqref{eq:theselemma5} is satisfied if $f_i(s)\geq0$,
$\forall s\in\{-1,1\}$. 
%
%
Applying Lemma~\ref{lem:equiv_alphag_mu}
again, we obtain 
\begin{align}
f_i(s)= \alpha_{g} x_{g+1}^{(j)} -& \mu_{g} \sum_{\l\notin \Qc\cup\{g+1\}} x^{(j)}_l\nonumber\\
    &-s\left(\alpha_{g} x_{i}^{(j)} - \mu_{g}
    \sum_{l\notin \Qc\cup\{i\}} x_l^{(j)} \right).
\end{align}
This leads to 
\begin{align}\label{eq:l5ineqf1}
f_i(1)=(\alpha_{g}+\mu_{g}) (x_{g+1}^{\text{${(j)}$}}-x_i^{\text{${(j)}$}})\geq 0,
\end{align}
since  $x_{g+1}^{(j)}\geq x_i^{(j)}\geq 0$ from \eqref{eq:h2}. 
Moreover, 
\begin{equation}
f_i(-1)
= (\alpha_{g}+\mu_{g}) x_i^{(j)}+(\alpha_{g}-\mu_{g}) x^{(j)}_{g+1} -
2 \mu_g\sum_{l\notin\Qc\cup\{g+1\}} x_l^{(j)}.\label{eq:fidemoins1}
 \end{equation}
 Hence, 
 from \eqref{eq:keyfeatx}
and $x_i^{(j)}\geq0$,
we have that $f_i(s=-1)$ is non-negative, and strictly positive 
when $g<j-1$.

 Finally, we obtain that $g+1$ is the unique minimizer of \eqref{eq:lemma5main} for $g<j-1$ by noticing that $x_{g+1}^{(j)}>x_i^{(j)}$ $\forall i>g+1$,  and that $\alpha_g>0$ and $\mu_g>0$ for $\mu<\frac{1}{g}$. The inequality in \eqref{eq:l5ineqf1} is therefore strict for $g<j-1$.
\hfill$\square$\\

We are now ready to proceed with the proof of the converse of Theorem
\ref{th:mainresult2}.\\

\textit{Proof (converse of Theorem \ref{th:mainresult2})}: We consider
the dictionary defined above. We assume that $\Qcs=\{1,\ldots,k\}$ and
$\Qcsc=\{k+1\}$.
 

Let us first show that \eqref{eq:cond1} is a universal bound for Oxx,
\ie it cannot be improved irrespective of the decay of the
coefficients. By setting $\mu=\frac{1}{k}$, we show that Oxx
leads  to a failure
during the first $k$ steps for any sparse representation 
$\y=\A\x$ supported by $\Qcs$. Indeed, suppose that Oxx has
selected atoms in $\Qc\subsetneq\Qcs$ during the first $k-1$ steps (if
not, the result trivially holds). Then, 
 denoting by
$\Qcs\backslash\Qc=\{i\}$ the remaining element in $\Qcs$,
we  show that
\begin{align}
|\langle \tc_i,\r^\Qc \rangle | &=  |\langle \tc_{k+1},\r^\Qc \rangle |, \label{eq:equalitysp}
\end{align}
\ie a failure situation such as \eqref{eq:OxxAmbiguityFailure} occurs.

From the
definition of the residual, we have $\r^\Qc=x_i \ta_i$. Since
  $x_i\ne 0$, proving \eqref{eq:equalitysp} is equivalent to
  showing that 
\begin{align}\nonumber
\vert\langle \tc_i, \ta_{i}\rangle \vert
=\vert\langle \tc_{k+1}, \ta_{i}\rangle \vert,
\end{align}
or, by using  Lemma \ref{lem:equiv_alphag_mu}, 
$\alpha_{k-1}=\mu_{k-1}$.
The latter equality can indeed be seen to be true 
 according to \eqref{eq:mudef_simpl}. We note that this result does
not depend on a particular instance of $k$-sparse vector $\x$ (and
thus on the decay) but only on the structure of the dictionary.



We now concentrate on the tightness of \eqref{eq:cond2}. Consider
the specific dictionary described above and the $k$-sparse vector
$\x^{(j)}$ defined in
\eqref{eq:thevampireslayer} with
$\mu=\frac{1}{2k- j}$, $j\in\{1,\ldots,k-1\}$. 
 It is easy to verify that the
 inequality in \eqref{eq:cond2}  is met for $i\neq j$, but becomes an equality for $i=j$. 
%
 It directly follows from
Lemma \ref{lem:goodatomselecCN} that
 our working hypothesis \eqref{eq:h2} is
  satisfied, and that $\Qc=\{1,\ldots,j-1\}$ is
selected during the first $j-1$ iterations. 
By virtue of the same lemma, we have
\begin{align}
\max_{ i\notin\Qc} |\langle \tc_i, \r^\Qc \rangle|=|\langle
\tc_j, \r^\Qc \rangle|
\label{eq:preuveconv2}
\end{align}
at the $j$th iteration ($g=j-1$). Moreover, using 
\eqref{eq:fidemoins1} and \eqref{eq:keyfeatx}
with $i=k+1$ and $g=j-1$ yields
$f_i(s=-1)=0$, \ie 
 $\langle \tc_{j}, \r^\Qc\rangle+\langle \tc_{k+1}, \r^\Qc\rangle=0$.
 We conclude
from~\eqref{eq:preuveconv2} that the failure situation
$\vert\langle \tc_{k+1}, \r^\Qc\rangle \vert =\vert\langle \tc_{j}, \r^\Qc\rangle \vert $ occurs.
\hfill$\square$

\section{Conclusions}
In this paper, we derived new guarantees of success for OMP and
OLS. First, we showed that there exists a sufficient condition of
success taking the form $\mu<\mu^\star$ with
$\mu^\star\in(\frac{1}{2k-1}, \frac{1}{k}]$ as soon as the nonzero
coefficients are not all equal (Theorem
\ref{eq:relaxation_theorem}). This result thus shows that the
traditional condition $\mu<1/(2k-1)$ can be weakened for decaying
vectors.  We then presented a new ``horizon-1" condition of success
taking the decay of the nonzero coefficients into account (Theorem
\ref{th:mainresult2}). In this condition, the specific upper bound
$\mu^\star$ is related to the rate of decay: the faster the decay, the
larger $\mu^\star$.  This condition reduces to $\mu^\star=1/(2k-1)$ as
soon as the two largest amplitudes in the sparse representation are
equal ($|x_1|=|x_2|$). Because the sparse vector may not be flat in
this situation, there is still room for further improvements. We note
however that generalizing ``horizon-1" conditions to more involved
settings may not be an easy task. 


Our decay-aware analysis of OMP/OLS also allowed us to carry out a
finer analysis of these procedures at intermediate iterations. We
considered both cases of ``partial support recovery" and ``successful
termination" using a non-empty initial support. For these two cases, we showed that the resulting conditions of success can be improved with respect to the standard $k$-step analysis (Theorem \ref{th:mainresult3}).

In the compressible and noisy case (for bounded-noise), we extended our $k$-step analysis 
by showing that the constraint on the mutual coherence is a function
of the noise amplitude, the compressible part of the sparse vector and the coefficient decay (Theorems \ref{corr:SRCnoisy} and \ref{th:noisymainresult}). One of our
results improves over the conditions proposed by Donoho \etal for OMP
\cite{Donoho2006Stable} when the sparse vector obeys some decay.
  
In the noiseless setting, we proved the tightness of the proposed
conditions (converse part of Theorem \ref{th:mainresult2}). First, we
emphasized that $\mu<1/k$ is a fundamental limit which cannot be 
improved.  Moreover, we showed that the proposed conditions of success are the best achievable guarantees of this type. To some extent, our results thus provide  a closure to the question of the characterization of the worst-case performance of OMP/OLS (in the noiseless setting)  as a function of the mutual coherence of the dictionary. 

Finally, let us mention that coherence-based conditions are easy to
evaluate and thus, of practical interest to have some insights into
the behavior of OMP/OLS. On the other side, coherence usually leads to
quite pessimistic conditions. In particular, the conditions proposed
in this paper, although improving over the standard uniform
guarantees, still require a number of measurements scaling as $m\sim
k^2$ to be satisfied for large underdetermined systems ($n\gg
m$).\footnote{This can be seen by invoking the Welch bound $\mu\geq
  \sqrt{\frac{n-m}{m (n-1)}}$, see \eg \cite[Th
  5.7]{FoucartMathematical}.} We note that other lines of work have
shown (in the uniform setting) that OMP/OLS can succeed in the regime
$m\sim k$, see \eg \cite{Foucart2013Stability, Zhang2011Sparse}. The
conditions derived in these works are however based on restricted
isometry constants and of practical interest only for large random
dictionaries since the explicit evaluation of restricted isometry
constants is an NP-hard problem \cite{Bandeira13, Tillmann2014Computational}.



\small
\bibliographystyle{IEEEbib}
\bibliography{group-15302_mod2}

\begin{IEEEbiography}{C\'edric Herzet}
 was born in Verviers, Belgium in 1978. He received the Electrical
Engineering degree and the Ph.D. degree in Applied Science from the
Universit\'e catholique de Louvain (UCL), Louvain-la-Neuve, Belgium, respectively
in 2001 and 2006. From May 2006 to December 2007, he was a post-doctoral
researcher within the ``Ecole Normale Supérieure de Cachan", Paris, and
the University of California, Berkeley (Fulbright scholarship). He is currently a
researcher within the  ``Institut national de recherche en informatique et automatique"
(INRIA), Rennes, France. His topics of research include inverse problems, low-rank
approximations and sparse representation algorithms.
\end{IEEEbiography}

\begin{IEEEbiography}{Ang\'elique Dr\'emeau}
was born in France in 1982. She
received the State Engineering
degree from T\'el\'ecom Bretagne, Brest, France,
in 2007 and the M.Sc. and the Ph.D. degree in
signal processing and telecommunications from the Universit\'e de Rennes, Rennes, France, in 2007 and 2010, respectively. She is currently an associate professor at ENSTA Bretagne. Her research interests include inverse problems, sparse representation algorithms and underwater acoustics.
\end{IEEEbiography}

\begin{IEEEbiography}{Charles Soussen}
(M'12) was born in France in 1972. He graduated from the \'Ecole
  Nationale Sup\'erieure en Informatique et Math\'ematiques
  Appliqu\'ees, Grenoble, France in 1996. He received the Ph.D. degree
  in Physics from the Universit\'e de Paris-Sud, Orsay, France, in
  2000 and his Habilitation \`a Diriger des Recherches in signal
  processing from the Universit\'e de Lorraine, France, in 2013. He is
  currently an Associate Professor at Universit\'e de Lorraine. He has
  been with the Centre de Recherche en Automatique de Nancy since
  2005. His research interests are in inverse problems and sparse
  approximation.
\end{IEEEbiography}

\end{document}